\documentclass[letterpaper, 10 pt, conference]{ieeeconf}  
\usepackage{blindtext}
\usepackage{graphicx}
\usepackage{color}
\usepackage{gensymb}
\IEEEoverridecommandlockouts                              
\title{\LARGE \bf
EIQIS: Toward an Event-Oriented Indexable and Queryable \\Intelligent Surveillance System}

\author{Seyed Yahya Nikouei$^{1}$, Yu Chen$^{1}$, Alexander Aved$^{2}$, Erik Blasch$^{2}$
\thanks{$^{1}$S. Y. Nikouei and Y. Chen are with the Dept. of Electrical and Computer Engineering, Binghamton University, SUNY, Binghamton, NY 13902, USA, Email: \{snikoue1, ychen\}@binghamton.edu}
\thanks{$^{2}$A. Aved and E. Blasch are with The U.S. Air Force Research Laboratory, Rome, NY 13441, USA, Email:
\{alexander.aved, erik.blasch\}@us.af.mil}
}

\begin{document}

\maketitle
\thispagestyle{empty}
\pagestyle{empty}

\begin{abstract}

Edge computing provides the ability to link distributor users for multimedia content, while retaining the power of significant data storage and access at a centralized computer. Two requirements of significance include: what information show be processed at the edge and how the content should be stored. Answers to these questions require a combination of query-based search, access, and response as well as indexed-based processing, storage, and distribution. A measure of intelligence is not what is known, but is recalled; hence, future edge intelligence must provide recalled information for dynamic response. In this paper, a novel event-oriented indexable and queryable intelligent surveillance (EIQIS) system is introduced leveraging the on-site edge devices to collect the information sensed in format of frames and extracts useful features to enhance situation awareness.
The design principles are discussed and a preliminary proof-of-concept prototype is built that validated the feasibility of the proposed idea. 

\end{abstract}

\section{INTRODUCTION}

Advances in networking, intelligence, and media available in urban areas attracts people towards a more comfortable lifestyle. Urbanization at an unprecedented scale and speed incurs significant challenges to city administrators, urban planners and policy makers. In order to efficiently manage the cities functions and be responsive to dynamic transitions, surveillance systems are essential for situational awareness (SAW) \cite{liu2014adaptive}, \cite{wu2015pseudo}. Nowadays, a prohibitively large amount of surveillance data is being generated every second by ubiquitously distributed video sensors. For example, North America alone has more than 62 million cameras in the year 2016. These cameras are connected to powerful data centers through communication networks and the delivery of surveillance video streams creates a heavy burden on the network. Researchers have shown that video streaming accounts for 74\% of the total online traffic in 2017 \cite{chen2017enabling}.

Since the first generation video surveillance systems known as Close Circuit TV (CCTV) were introduced in 1960s, urban surveillance mechanisms adapted to the changing technology \cite{surette2005thinking}. Compared with today’s edge computing paradigm, CCTV-like surveillance systems are limited because:

\begin{itemize}

   \item The network is ``best effort'' based which means not only transmission of the video data suffers delays and jitters, the data may get lost or dropped because of network congestion. 
   \item The raw-data transmission is ``dedicated'' which wastes resources in the communication network and at the data center, because not all data is globally significant or worthy to be stored for long time.   
   \item An agent needs to pay “full attention” to the video to capture any emergency in real-time. Obviously this naïve approach is not scalable, and there are several architectures introduced based on computer vision techniques and make decisions based on machine learning algorithms. However, to date there is not a system that is able to meet the performance requirements like real-time, good scalability, and robustness  \cite{tsakanikas2017video}. 
   \item An agent employs ``working memory'' as computing capabilities afforded only searching for a specific target of interest or focusing on a special feature. Meanwhile, today’s multimedia forensics desires real-time or near real-time searching by scanning through the large surveillance video record base.

\end{itemize}

It is very challenging to immediately analyze the objects of interest or zoom in on suspicious actions from thousands of video frames. Making the big data indexable is critical to tackle the object analytics problem \cite{aved2015multi}, \cite{blasch2015dynamic}. It is ideal to generate pattern indexes in a real-time, on-site manner on the video streaming instead of depending on the batch processing at the cloud centers. The modern edge-fog-cloud computing paradigm allows implementation of time sensitive tasks at the network edge. In this paper, a novel event-oriented indexable and queryable intelligent surveillance (EIQIS) system is introduced leveraging the on-site edge devices to collect the information sensed in format of frames and extracts useful features to enhance situation awareness. 

The rest of this paper is organized as follows. Section 2, briefly discusses background knowledge and relative work. Section 3 highlights the main challenges in the real-time surveillance. Section 4 introduces the rationale of the proposed indexable and queryable surveillance system. A preliminary study is presented in Section 5, which validates the concept and shows the feasibility of the system architecture. Finally, Section 6 concludes the paper with future research directions.

\section{Background Knowledge and Related Work}

Today, most available surveillance systems archive streaming video footage to be used off-line for forensics analysis  \cite{chen2018smart}. Communication delays and uncertainties associated with the data transfer from image sensors to a remote computing facility limit implementation of the online surveillance tasks. However, delay sensitive applications require on-line processing. Thanks to the recent development of lightweight machine learning (ML) algorithms that require less computing power and storage space, more processing can be migrated to the edge of the network \cite{ouaddah2016fairaccess}, where no more delay is incurred for data transmission. For tasks like anomalous behavior detection that is not affordable at the edge, instead of directly outsourcing the job to the remote cloud, near-site fog nodes are powerful enough for complex data analytics tasks. 

For instance, in a smart transportation application following a hierarchical system architecture, data is accessed by the sensors implemented on buses and transferred to a fog node where contextualization and decision making happens \cite{chamasemani2013systematic}. For video surveillance systems, the remote cloud is mainly used for profile building, pattern analysis, and long term historical record analysis. 

\begin{figure}[t]
    \centering
    \includegraphics[width=0.45\textwidth]{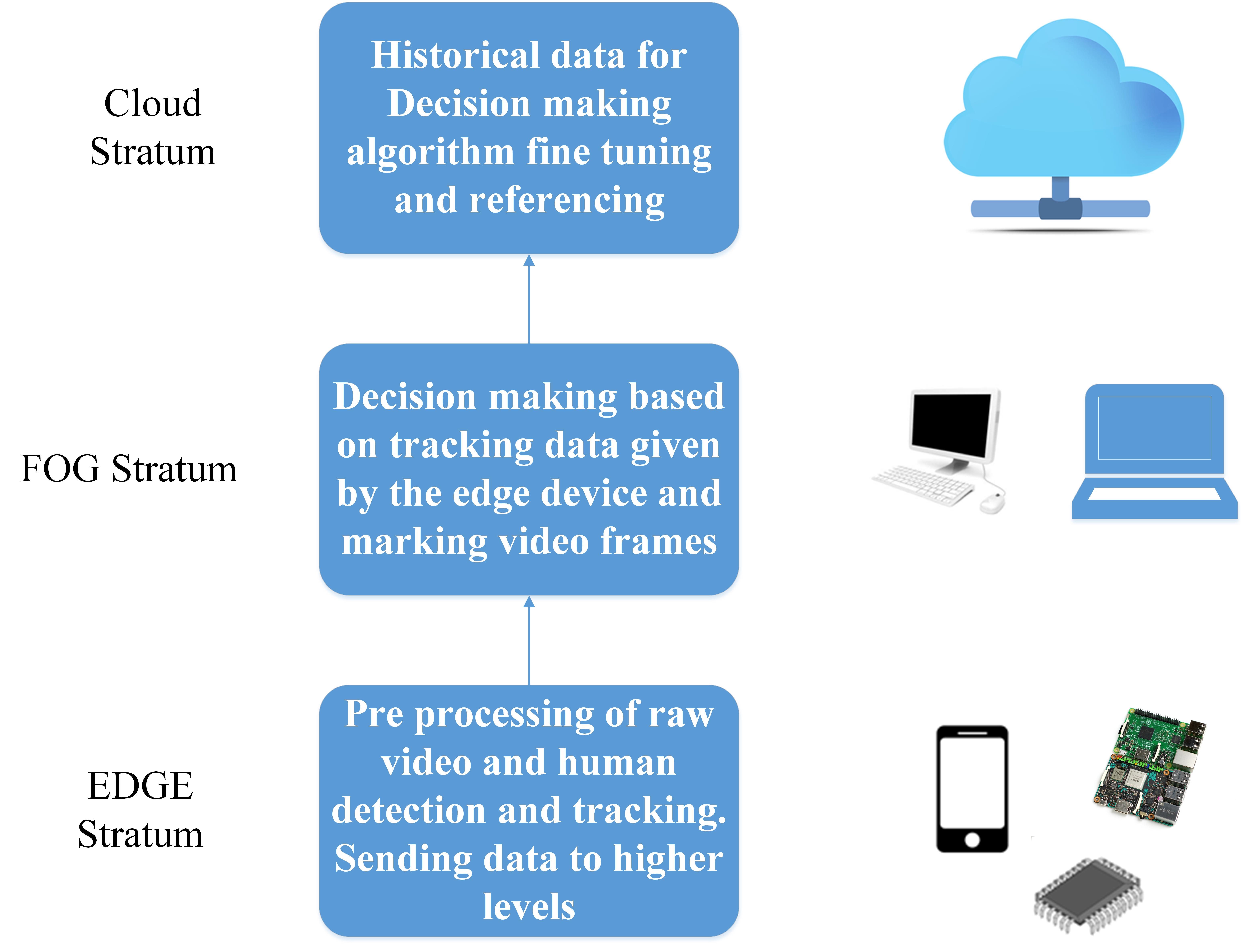}
    \caption{Layered smart surveillance system hierarchy using the edge-fog-cloud computing paradigm.}
    \label{fig:arch}
    \vspace{-10pt}
\end{figure}

In general, a smart surveillance system includes three layers as shown in Fig. \ref{fig:arch}. In the first layer, image analysis, the input camera frame is given to an edge device and the low-level features are extracted \cite{nikouei2018real}, \cite{penmetsa2014autonomous}. The edge devices are able to conduct object detection and object tracking tasks \cite{khanezaei2014framework}, \cite{yu2018survey}. The intermediate-level, considered as the fog stratum, is in charge of mode recognition for action recognition, behavior understanding, and abnormal event detection. Finally, the high-level, cloud center, is focused on systems analysis including historical profile building, global statistical analysis, and narrative reporting. Connections among the edge, fog and cloud nodes present challenges in terms of overall platform, connections, quality of service (QoS) requirements, and preserving privacy and security.

The first step of a video surveillance system is to simultaneously track and identify (ID) (STID) the objects of interest in the video \cite{blasch2005multiresolution}, \cite{blasch2000data}. STID continues to be a challenging task performed on the edge of the network \cite{nikouei2018intelligent}. Nowadays, once an event incurred, the operators need to spend considerable amount of time to go through the footage and look at videos from different cameras in order to find a specific target. Even in the next generation surveillance systems that are combined with image processing techniques for better decision making, performing a search in real-time or near real-time is very challenging \cite{blasch2014context}, \cite{tsakanikas2017video}.  

Ideally, the surveillance system is expected to be able to quickly and automatically identify the clips of interest based on a given query. Earlier researchers have proposed to adopt video parsing techniques that automatically extract index data from video and store the index data in relational tables \cite{blasch2014quest}, \cite{hammoud2014automatic}, \cite{hampapur2007searching}. The index is used through SQL queries to retrieve events of interest quickly. However, this approach cannot meet the performance requirements of online, real-time, operator-in-loop interactions. Future smart surveillance video streams have to be indexable and queryable such that the operator is able to obtain the information of interest instantly.

\section{Real-time Queryable Surveillance: Architecture and Challenges}

This section introduces an edge-fog-cloud computing based system architecture to achieve event-based indexable and queryable intelligent surveillance (EIQIS). It is non-trivial to extract features in real-time and use them as indexes to conduct online query on surveillance video streams \cite{palaniappan2010efficient}. Advances in machine learning, multi-modal data fusion, and physics-based and human-derived information fusion (PHIF) show promise for EIQIS. Current systems are designed to enhance user responsibilities to include security, surveillance, and forensics. Typically, the user provides a standing query that the image processing is to provide event triggers \cite{aved2015multi}, \cite{blasch2015urref}.  The user would like the system to do the functions autonomously, however, the ultimate design would include a combination of humans in, on, or out-of the loop (HIL, HON, HOON).

In order to have a smart surveillance system raise an alarm when something abnormal is detected, each captured frame that is processed requires knowledge of the proceeding frames. A three layer edge-fog-cloud hierarchical architecture reduces the delays that are incurred when the frame is transferred to a remote cloud center. The more processing that is migrated to the network edge, the faster the features are obtained and indexes are constructed because of the close proximity of the edge node to the geo-location of the camera. Meanwhile, due to the constraints on computation and storage capacity at the edge devices, more computing or data intensive tasks are outsourced to more powerful cloud.

The first layers is the edge camera, it should be mentioned that most reliable detection and tracking algorithms are dedicated for specific surveillance applications. Running them in a resource constrained environment that requires the algorithm to be a light weight version of the original does not help the accuracy. Thus, finding better methods is a contemporary research topic \cite{li2017dynamic}.

Once a frame is captured by the image sensor, it will be either transferred to the edge device that is connected via a local area network (LAN) connection or processed on-site if it is a smart camera (edge device) with sufficient computing power. The edge node has limited computing power and so all computing intensive event detections cannot be executed at this level. The edge device conducts pre-processing using a convolutional neural network (CNN), which will identify the objects of interest and give their positions in the image frame. Even with small architectures with few layers that reduce the overall computation complexity, CNNs are heavy for the edge device  \cite{nikouei2018intelligent}. The edge device cannot afford to execute the CNNs more than couple of times per second. Therefore, in order to reach a higher resolution of the detection, the bounding box around the object of interest is given to a tracker algorithm that uses an online learning algorithm to follow the object in each frame until it moves out of the frame. Each time the CNN runs, the newly found bounding boxes are sent to a fast tracker such as the Kernelized Correlation Filter (KCF), improving the speed.   It should be noted that although newer and powerful edge nodes are made every day, with more features to be extracted, a longer processing time is needed. Consequently, the key for the real-time application is a trade-off between the speed and the amount of features to be extracted in each frame.

After each object is detected and tracked, features can be extracted. These features might include, but are not limited to the current position and speed the object is walking, the direction of the walk and some other physical features such as the angles the other parts of the upper body parts create and so the pose of the pedestrian \cite{turaga2008machine}. For each detected pedestrian, there is a table that is updated with each frame and includes a key and value for features extracted from the video. The actual video may not be needed to be transferred to the fog level device where the decision making code is executed. 

The edge device is designed to conduct immediate techniques such as feature extraction, while the advanced analytics is outsourced to a more powerful, near-site node. Several edge devices from several camera feeds can be connected to a fog node, which conducts feature contextualization, indexing, and storage. One of the challenges in a surveillance system is the security of the connection between the edge and fog. Although there are new promising technologies to address privacy/security, like blockchain technology \cite{xu2018blendcac}, more development is needed to make them light weight and robust for the smaller networks with low power. The features transmitted to the fog node can be contextualized to support decision making \cite{snidaro2016context}. Valuable data in the contextualization include: The location of the camera, time of the footage, terrain information, semantic ontologies of descriptors, etc. For example, while it is normal for people to walk and stand in a campus building, it can be considered as abnormal if it is late at night when the building should be close. Also, connecting several cameras in the same area to the same fog node will give the fog the ability to look at the monitored area from different perspectives, illuminations, and contexts. 

Another challenge that the surveillance community faces is the decision support algorithm, which includes supervised, unsupervised, and semi-supervised methods. The, the lack of labeled data for unknown situations, requires methods in semi-supervised training to better characterize abnormal situations. The answer may include the location and several other factors and sequence of events lead to abnormal behavior detection. Also, the security camera and the functionality of the place surveillance may differ from one to the other which makes it very difficult to differentiate between normal and abnormal activity.

The historical analysis, profile building, and situation analysis are conducted by the most powerful node in the edge-fog-cloud architecture hierarchy, the cloud. The decisions making and the detection of false alarm and the features that raised the alarm are sent for future fine tuning of the algorithms and also some analytical studies. Figure 1 shows the interconnections of the nodes in the network described in this section.

\section{Making the Video Streams Indexable}

The usability of any exploited video is based on what is stored or indexed or fast retrieval, such as content-based image retrieval. The surveillance video streamed to the edge device enables features extracting for decision making. Decision making is based on the real-time search query. The real-time video search will make the job of the operator/user easier by giving instances of the video that are asked for in a query to the system. Search string is the query that is given to the fog node. The fog node is the ideal level to handle search requests where contextualized information from close by cameras is stored. The following describes how a query is handled at the fog layer:

\begin{enumerate}

   \item The fog node receives the query and will check the eligibility of the machine asking for the information. The access level of the nodes in such a network is defined in a smart contract in a blockchain enabled security platform. 
   
   \item The fog node searches for the query in the index table to find the corresponding camera, timestamp and other information based on the real-time features provided and select them if any.

   \item The fog node answers the search requester based on the information found. 

   \item Then the operator selects the cameras with the query and has the live feed or recorded clips (it is assumed that the operator has access to the edge device in charge of the camera of interest if he/she has access to the higher-level fog).
\end{enumerate}

The operator thus can search the video streams in real-time. 

Indexing requires the association of complementary information (hashed, correlated, and linked) with the video frame for storage. Using the mapping table affords fast information retrieval. Considering the indexing table the same as the features simplifies the search operation. While there are many features extracted from the video, there might be several different indexes that are required by the system administrator. Features are generated in order to make a decision for the actions of the object in the video. However, indexes that are based on features might include more options. There are two scenarios that are plausible. First, the fog node uses the same features and adds context to make the data useable as the index table. Second, the fog node uses several edge devices (perform as microservices) to extract features required and creating a table to be used as indexes based on the resulting features.

\subsection{Indexing}

In order to facilitate faster search results, one known method used today in search engines and operating systems, is to create an index table which is used later for finding search queries. Indexing means to have a key and value table of features that are of interest and once the keys are searched for (in query format), the corresponding values are the results of the search that gives certain files that contain the query. This way the search is faster and there is no need to scan all files for the key values that are searched for. The same principle applied to the video file captured by the surveillance cameras results in efficient and real time operations. Based on the index table points to the corresponding edge device, the camera live or recorded footage clips are identified and sent to the query sender.

Once the camera captures each frame, an edge device extracts features in real-time or near real-time from the video and the features are transferred to a fog node. After the contextualization of the features, they can be used as the indexes for querying when the operator needs to find something instantly. For example, if the operator is looking for moments that there are congestion of people on the campus in the late night hours. The search can be directed to the exact hours and locations, then look for features that report more than ten people or more at the same frame. Using the query-based parameters inherent in the index table will lead to the corresponding video clips faster and the operator can look for incidents that have the exact search keys. The EIQIS method is obviously more efficient than having to check all the camera footage security systems to find what imagery is of interest.

\subsection{Features VS. Indexes}

\begin{figure}[t]
    \centering
        \includegraphics[width=0.425\textwidth]{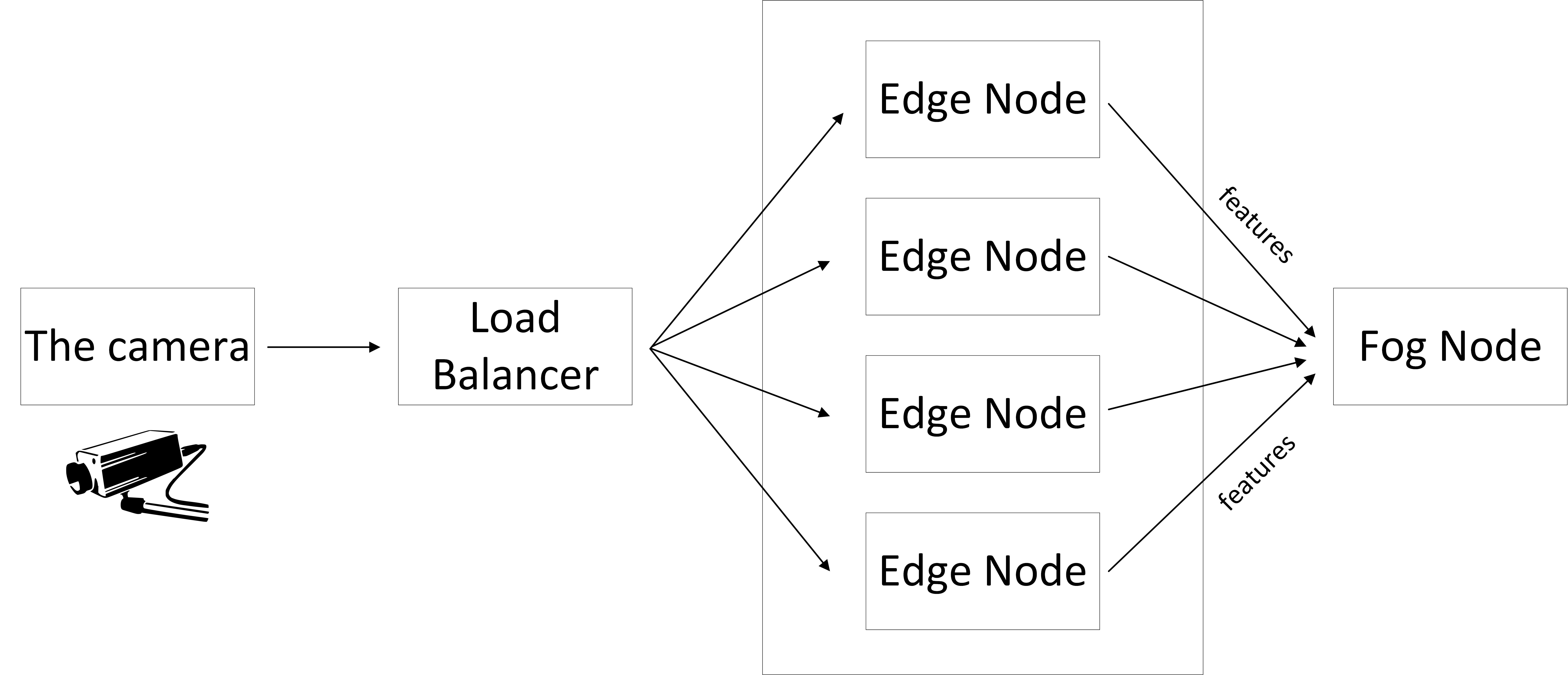}
    \caption{Edge feature extraction as microservices for indexing purposes.}
    \label{fig:fig_2}
    \vspace{-10pt}
\end{figure}

Creating the indexes for the extract features that are useful for video search supports historical analytics. However, the features that are of interest in the abnormal behavior detection may not support an operator search, be enough, or exactly the same as the indexes (key values) that are applicable in usual search. Figure \ref{fig:fig_2} shows a scenario in which more feature extraction from the video is needed. The job can be divided into more than one edge devices and each feature can be handled as a microservice \cite{nagothu2018microservice}. Microservices is defined as a separate piece of program that provides a service to a bigger piece of program. In this case the feature extraction can be considered as the microservice that is used in the video indexing platform. More features can be extracted as a result of this architecture. If any indexes need to be added, simply adding the service to the platform can expand the scope of the indexes that are used. 

\section{A Preliminary Case Study}

\begin{figure}[t]
    \centering
        \includegraphics[width=0.425\textwidth]{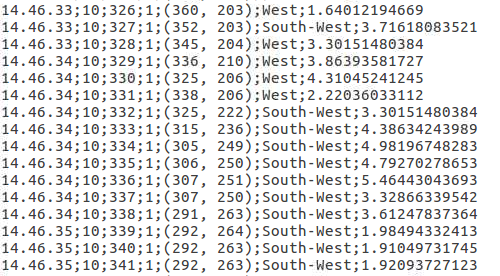}
    \caption{Example feature table for each camera.}
    \label{fig:fig_3}
    \vspace{-10pt}
\end{figure}

\begin{figure}[t]
    \centering
        \includegraphics[width=0.425\textwidth]{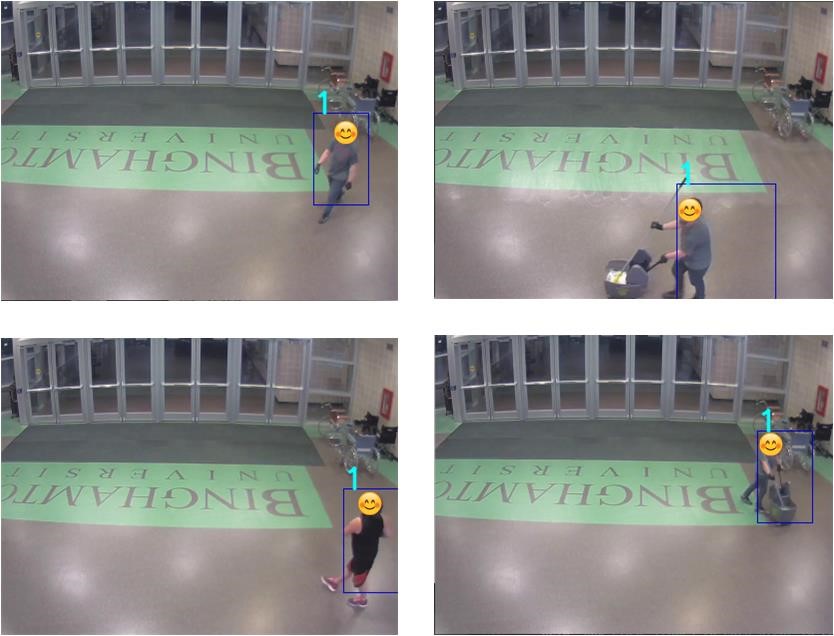}
    \caption{Viualized features in real-time.}
    \label{fig:fig_4}
    \vspace{-10pt}
\end{figure}

A preliminary proof-of-concept prototype has been built to validate the feasibility of EIQIS \cite{nikouei2018realb}. It shows that the edge devices are capable of extracting and sending features in real-time to the fog layer. The features are written into a text file and sent to fog through a secure channel. The features are synchronized with every node of the network for added security. Figure \ref{fig:fig_3} is an example of features stored in the fog in a key value manner and Fig. \ref{fig:fig_4} is graphical output of the edge device, where the device adds a bounding box around the object (e.g., person, vehicle, other) of interest and the box follows the object. Figure \ref{fig:fig_4} presents several moments that are challenging to be detected. It is a proof showing an acceptable performance of the edge device. 

The real environment validates the feasibility of the proposed system. The prototype model run on two Asus Tinker Boards with the configuration as follows: 1.8 GHz 32-bit quad-core ARM Cortex-A17 CPU, the memory is 2GB of LPDDR3 dual-channel memory and the operating system is the TinkerOS based on the Linux kernel. The fog layer functions are implemented on a laptop, in which the configuration is as follows: the processor is 2.3 GHz Intel Core i7 (8 cores), the RAM memory is 16 GB and the operating system is the Ubuntu 16.04. A private blockchain network is implemented to secure the feature data transferring from edge to fog. Our private Ethereum network includes four miners, which are distributed to four desktops that are empowered with the Ubuntu 16.04 OS, 3 GHz Intel Core TM (2 cores) processor and 4 GB memory. Each miner uses two CPU cores for mining task to maintain the private blockchain network and the resulting blocks are synchronized through the whole network so every node has a copy of the latest block. The data transfer between the fog node and the miner is carried through an encrypted channel. Before the fog node can secure the features, there should be no adversaries who can temper with the surveillance data. Python based socket programming language is used for both ends of the channel. More details of the prototype are reported in \cite{nikouei2018realb}.

\section{CONCLUSIONS}

Many surveillance systems available today cannot meet the performance requirements raised from real-time, human-in-loop interactive operations. The event-oriented indexable and queryable intelligent surveillance (EIQIS) edge-fog-cloud hierarchical architecture is promising for real-time or near real-time applications, which allows instant querying on the online surveillance video streams to give more time to first responders. In this paper, the architecture toward an event-oriented, indexable, queryable smart surveillance system is introduced. The proposed system enables query of video in real-time based on an index table, which is created on top of the features that are extracted on-site by edge computing nodes. This intelligent surveillance system enables the operator to search for scenes or events of interest instantly. A preliminary study has validated the feasibility of the proposed architecture.  

\addtolength{\textheight}{-12cm}   








\bibliographystyle{IEEEtranS}

\bibliography{MMTC.bib}

\end{document}